\documentclass[12pt,epsf,epsfig,psfig]{article}
\usepackage{graphicx}
\usepackage{epsfig}
\usepackage{cite}
\def\J{$J/\psi$}

\def\X{$\chi_c$}

\def\P{$\psi'$}

\def\C{c{\bar c}}

\def\NP{{ Nucl.\ Phys.\ }}
\def\PL{{ Phys.\ Lett.\ }}
\def\PR{{ Phys.\ Rev.\ }}

\def\PRL{{ Phys.\ Rev.\ Lett.\ }}

\def\ZP{{ Z.\ Phys.\ }}

\def\be{\begin{equation}}
\def\ee{\end{equation}}
\def\lsim{\raise0.3ex\hbox{$<$\kern-0.75em\raise-1.1ex\hbox{$\sim$}}}
\def\gsim{\raise0.3ex\hbox{$>$\kern-0.75em\raise-1.1ex\hbox{$\sim$}}}

\begin{document}

\parindent=0pt 

~ \hfill BI-TP 2011/41

\vskip1.5cm

\centerline{\Large \bf Charmonium Production and Corona Effect}

\vskip1cm

\centerline{\bf S.\ Digal$^{1,2}$, H.\ Satz$^2$ and R.\ Vogt$^{3,4}$}

\bigskip

\centerline{$^1$The Institute of Mathematical Sciences, Chennai, India}

\centerline{$^2$Fakult\"at f\"ur Physik, Universit\"at Bielefeld, 
D-33501 Bielefeld, Germany}

\centerline{$^3$Physics Division, 
Lawrence Livermore National Laboratory, Livermore, CA 94551, USA}

\centerline{$^4$Physics Department, University of California, Davis, 
CA 95616, USA}

\vskip1cm

\centerline{\bf Abstract}

\bigskip

We study the centrality dependence to be expected if only charmonium 
production in the corona survives in high energy nuclear collisions, 
with full suppression in the hot, deconfined core. 
To eliminate cold nuclear matter effects as far as possible, we consider
the ratio of charmonium to open charm production. The centrality 
dependence of this ratio is found to follow a universal geometric form, 
applicable to both RHIC and LHC in collisions at central 
and forward rapidities. 

\vskip1cm 

The behavior of charmonium production in high energy nuclear collisions was
proposed quite some time ago as a probe of color deconfinement \cite{M-S}. 
The basic idea was that color screening in the hot primary medium prevents
the binding of $\C$ pairs produced in the early stages of nucleon-nucleon
collisions. As a result, the ratio of hidden to open charm production 
should vanish when the energy densities of the medium produced
in the collision are sufficiently high. Since the observed \J~production rate 
is partially due to feed-down from higher mass excited states, the suppression 
should occur in
a sequential fashion, first for the decay products from \P~and \X~production, 
then for directly produced \J($1S$) \cite{KMS,KS91,KKS}. 

\medskip

\begin{figure}[htb]
\centerline{\epsfig{file=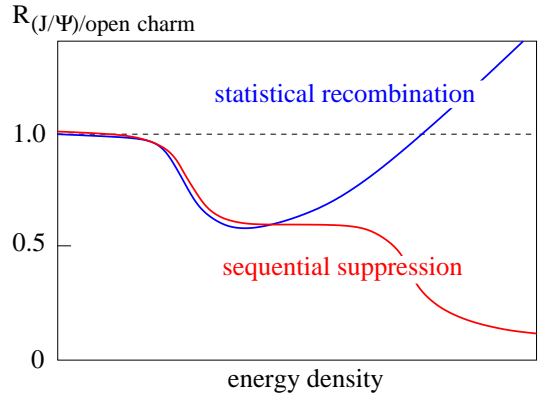,width=7cm}}
\caption{The schematic ratio of \J~to open charm production in nuclear 
collisions, normalized to the scaled $pp$ value, 
as a function of the energy density.}
\label{comp1}
\end{figure}

In subsequent work \cite{PBM,Thews,Rapp}, it was argued that, at the
hadronisation point, statistical combination of the charm quarks present
in the deconfined medium could lead to regeneration of charmonia.
Even if the ``primary'' charmonium states produced in individual 
nucleon-nucleon
collisions are dissociated through color screening, the abundant
charm production in high energy collisions could provide enough 
charm quarks to result in ``secondary'' charmonium formation through the 
binding of combinatorial pairs
of $c$ and $\overline c$ quarks produced in different initial nucleon-nucleon
collisions. Since the charm production rate grows with collision energy
and centrality, this would imply a corresponding increase in the ratio of 
hidden to open charm. In 
Fig.~\ref{comp1}, we schematically compare the statistical recombination 
prediction to that of the sequential color screening scenario. The 
ratio of hidden to open charm in nuclear collisions is normalized 
to the $pp$ value, scaled by the number of binary nucleon-nucleon
collisions.  The increase of this ratio above unity in central collisions
for high charm production rates is a crucial prediction of the 
statistical recombination model. On the other hand, the sequential suppression
scenario predicts that the ratio should decrease to values well below the
central SPS results at high energy density when direct \J($1S$) suppression
sets in.

\medskip

\medskip

The first RHIC results on charmonium production, after a decade of SPS data,
showed a rather striking feature.  While it
seemed clear that the energy density of the produced medium should be much
higher at RHIC than at the SPS (typical values are 5 - 6 GeV/fm$^3$ at RHIC
relative to 2 - 3 GeV/fm$^3$ at the SPS), the magnitude of \J~suppression in 
central RHIC collisions was compatible with the SPS data within the measured 
uncertainties, as was the overall centrality dependence.  Both the sequential 
suppression and statistical recombination scenarios could explain this behavior
with different sets of assumptions:
\begin{itemize}
\vskip-0.2cm
\item{In the sequential suppression approach, full suppression of the excited 
\P and \X~states is expected, with complete survival of
the directly-produced \J($1S$) states, after cold nuclear matter effects are
accounted for;}
\vskip-0.2cm
\item{The statistical recombination approach predicts a monotonically 
decreasing survival probability for all $\C$ states, coupled with secondary 
regeneration increasing with energy density.} 
\end{itemize}
Predictions for the new data from the LHC in the two scenarios, at still higher 
energy density (at least 8 - 10 GeV/fm$^3$), lead to relatively distinct
outcomes: vanishing measured \J~yield with sequential suppression and 
considerably enhanced \J~production with secondary regeneration. 
In both cases, the predictions 
are cleanest for the ratio of charmonium to open charm since, 
in this case, initial-state effects (shadowing, parton energy loss) largely 
cancel \cite{sridhar}. Because such data are not yet available, the 
centrality dependence of the cold nuclear matter effects introduces some 
uncertainty.  However, the presently available data do not rule out the 
possibility that once all corrections (\J~feed-down from $B$-decay, 
comparable rapidity and transverse momentum ranges) are made, the \J~production 
rates at the LHC, relative to open charm production, will be quite comparable 
to the corresponding ratios in central RHIC and SPS collisions. 

\medskip

The aim of this note is to argue that, if indeed the fraction of charmonia 
survival in nuclear collisions is essentially
independent of the collision energy, then the survival probability at high
energy density appears to be
a geometric effect due to charmonium production in the corona of the collision
where the density is too low for dissociation of primary
charmonium production.

\medskip

\begin{figure}[htb]
\centerline{\psfig{file=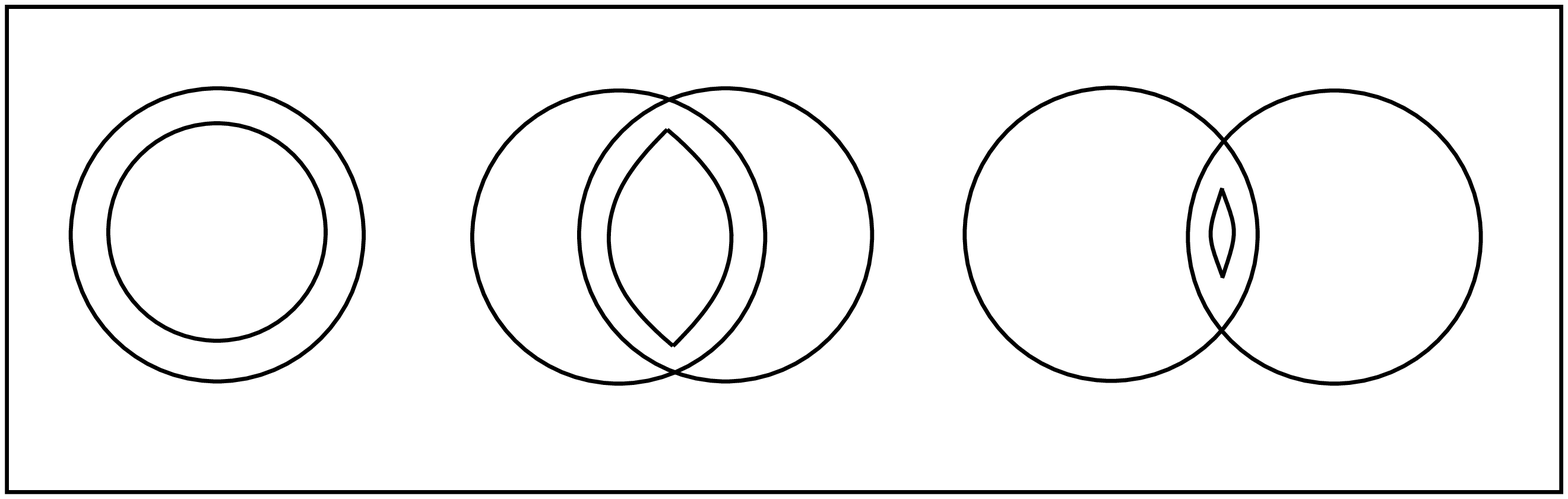,width=10cm}}
\caption{Corona contributions for decreasing centrality, as seen in the 
transverse profiles of the colliding nuclei.}
\label{cor}
\end{figure}

The so-called corona effect in nuclear collisions has been discussed in quite 
some detail for various observables
\cite{stock,beca,bozek,hohne,werner,beca-man,andronic,aichel}.
Its role in heavy quark production has also been considered
\cite{andronic}. 
The basic feature is that, even in a completely central collision, 
there is a rim region in the transverse plane
in which the density of wounded nucleons is low, or in which each 
nucleon undergoes very few collisions, perhaps only a single one. In this
region, there is no produced medium so that, in the corona, all observed effects
should be similar to those in $pp$ or $pA$ collisions. While
in central collisions of heavy nuclei, the ratio of the rim area to the total 
overlap area can be quite small (depending on how the corona region is defined),
the fraction of the collision region attributable to the corona increases with 
increasing peripherality until, in the most peripheral
collisions, there are effectively just single nucleon-nucleon interactions
(see Fig.~\ref{cor}). In the aforementioned studies
\cite{stock,beca,bozek,hohne,werner,beca-man,aichel,andronic}, 
the aim was typically to
remove or correct for the corona effect. Here we take the opposite
approach: we assume that the core completely suppresses all charmonium
production so that only the corona structure and the resulting centrality
distribution are relevant.

\medskip

\medskip

Before proceeding, we briefly recall the definitions of the relevant geometric
quantities. We use the formulation given in a recent survey \cite{Miller}, 
to which we refer for further details.
We consider a collision of two nuclei $A$ and $B$. For a nucleus of mass 
number $A$, the nucleon density projected onto a plane orthogonal to the
collision axis is given by 
\begin{equation}
{\cal T}_A({\bf s})=\int dz_A \rho_A({\bf s},z_A) \, \, .
\end{equation}
Here $\rho_A$ is the probability, normalized to unity, to find a nucleon 
from nucleus $A$ at the location $({\bf s},z_A)$, where $z_A$ denotes the
longitudinal position and ${\bf s}$ is the position in the transverse plane. 
For an impact parameter ${\bf b}$, 
the density of the number of collisions in the transverse plane is given by
\begin{equation}
n_c({\bf s},{\bf b})=AB{\cal T}_A({\bf s}){\cal T}_B({\bf s}-{\bf b})
\sigma^{NN}_{\rm inel} \, \, ,
\label{ncdef}
\end{equation} 
where $\sigma^{NN}_{\rm inel}$ is the nucleon-nucleon inelastic cross section.
The integral over ${\bf s}$ then gives the total number of collsions at impact
parameter ${\bf b}$. 
Similarly, the wounded nucleon density is given by
\begin{eqnarray}
n_w({\bf s},{\bf b})=A{\cal T}_A({\bf s})\left\{1-\left[1-{\cal T}_B
({\bf s}-{\bf b}) \sigma^{NN}_{\rm inel}\right]^B\right\}&& \\\nonumber
+B{\cal T}_B({\bf s}-{\bf b})\left\{1-\left[1-{\cal T}_A({\bf s})
\sigma^{NN}_{\rm inel}\right]^A\right\}&& \, \, .
\label{nwdef}
\end{eqnarray}
The integral of $n_w$ over ${\bf s}$, $N_{\rm part}({\bf b})=\int d^2s 
n_w({\bf s,b})$, gives the total 
number of wounded nucleons (or participants) for a given centrality ${\bf b}$. 

\medskip

\begin{figure}[htb]
\centerline{\psfig{file=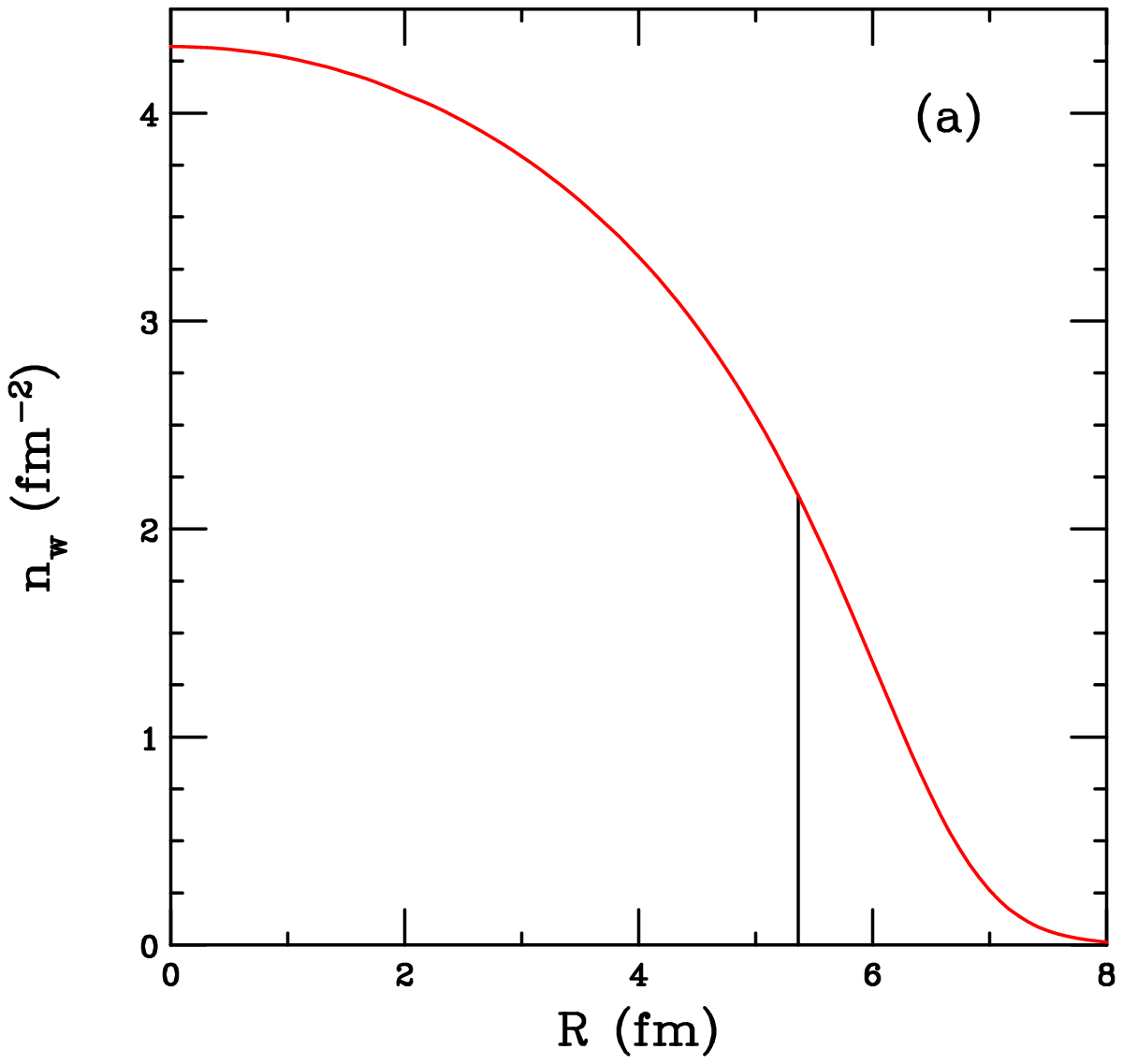,width=7cm}\hfill
\psfig{file=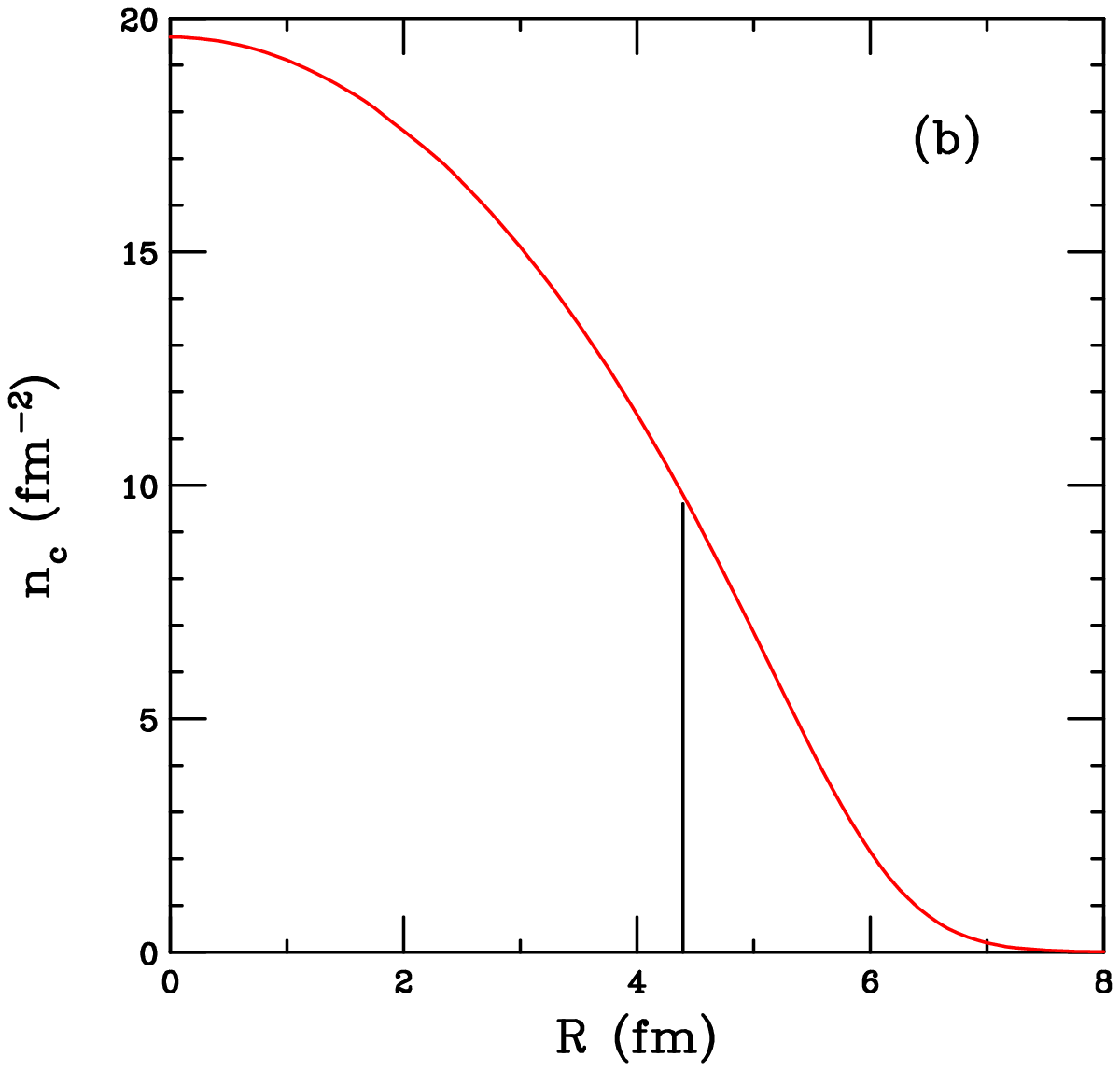,width=7cm}~~~~~~~}


\caption{Transverse profiles in central Au-Au collisions, showing
(a) the density of wounded nucleons, $n_w$, and (b) the density of 
collisions, $n_c$, in the transverse plane as a function of the radius $R$,
calculated with $\sigma_{\rm inel}^{NN} = 42$ mb. 
In both cases, the corona is defined as that region in which the density 
($n_w$ or $n_c$ respectively) has decreased by a factor of two or more.}
\label{prof}
\end{figure}

The ultimate magnitude of the ratio of contributions from the corona region to 
those from the overall interaction region depends both on the variables used 
to distinguish between the rim and the central core regions (density of wounded 
nucleons, collision density, energy density) and on the chosen thickness used,
{\it i.e.}, on the definition of the corona. Since the radius of a nucleus is
generally defined as that distance from the center at which the matter density
distribution, $\rho_A$, has decreased by a factor of two, we use a similar
definition to specify the edge of the corona in nuclear collisions. 
The resulting transverse profiles of the
density of wounded nucleons and of the collision density, both for central 
Au+Au collisions, are shown in Fig.~\ref{prof} for $\sigma_{\rm inel}^{NN} = 
42$~mb. For this system, 
$n_w = n_w(R=0)/2 = 2.16$~fm$^{-2}$ at $R = 5.4$~fm
while $n_c = n_c(R=0)/2 = 9.6$~fm$^{-2}$ at $R = 4.4$~fm.

\medskip

\begin{figure}[htb]
\centerline{\psfig{file=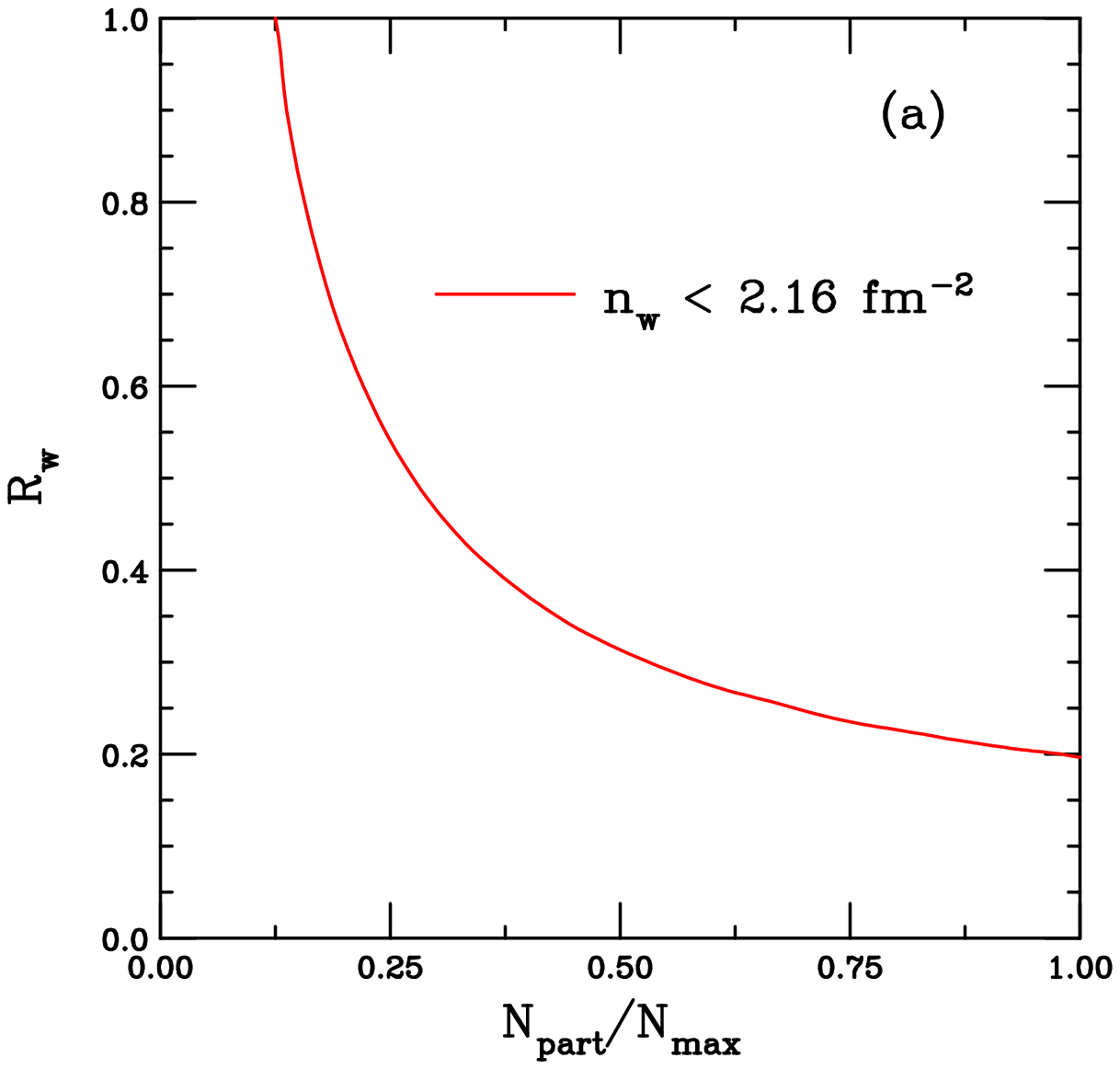,width=8cm}\hfill
\psfig{file=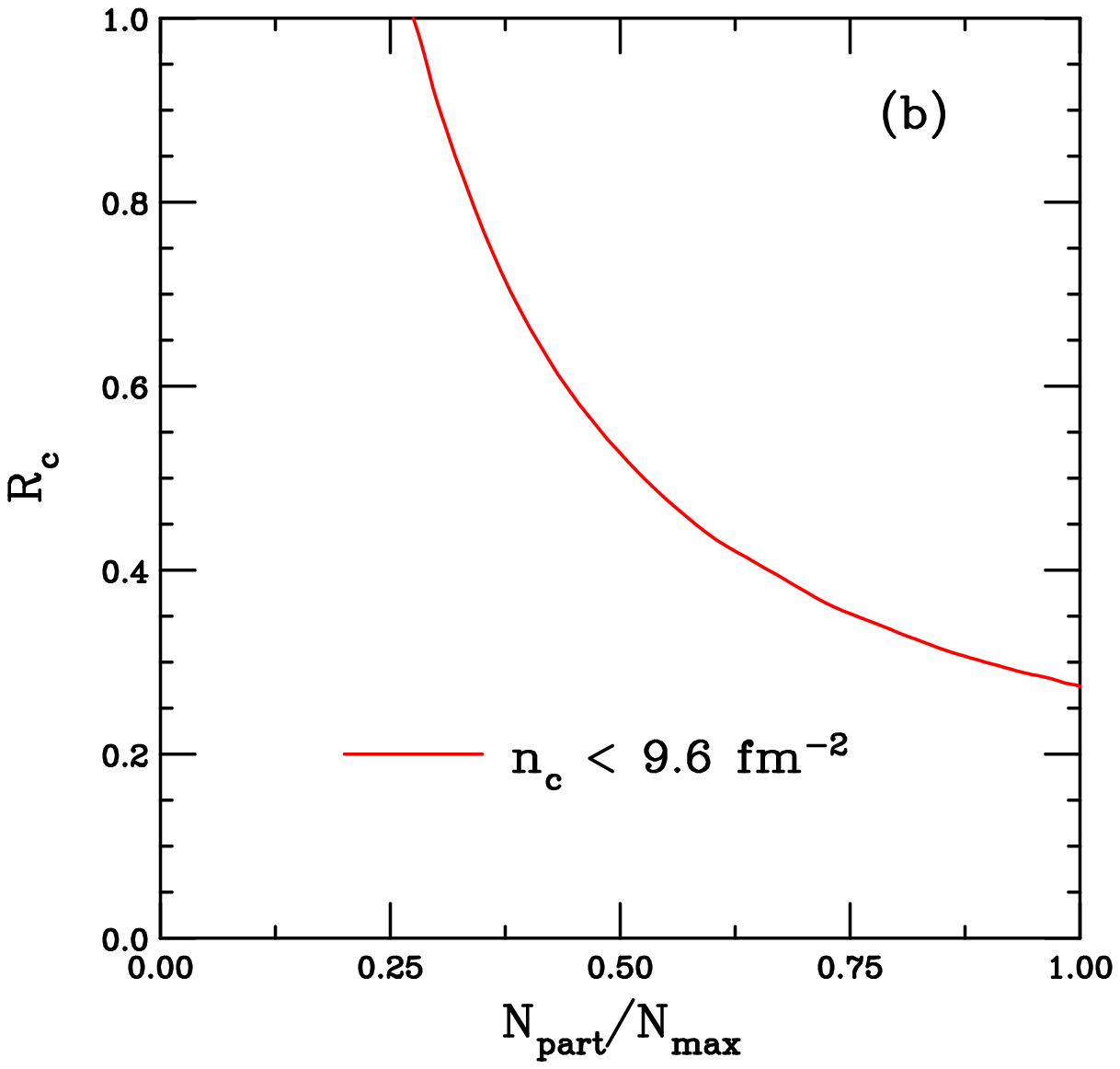,width=8cm}~~~~~~~}


\caption{(a) Ratio of wounded nucleons in the corona region to the total
number of wounded nucleons.  Here the corona region is defined as the region
where the density of wounded nucleons, $n_w$, has decreased by a factor of
two or more relative to the value at $b=0$, $n_w < 2.16$~fm$^{-2}$. (b)  Same
as (a) for collision density with $n_c < 9.6$~fm$^{-2}$.  Both calculations are
for Au+Au collisions with $\sigma^{NN}_{\rm inel} = 42$~mb.} 
\label{rim}
\end{figure}

A general feature emerges from the geometry of the collision: for a 
considerable range of impact parameters from central through mid-peripheral 
collisions, the fraction of the transerve collision area in the corona 
relative to the total transverse overlap is small and relatively constant. 
Only in very peripheral collisions, roughly corresponding to $b > 1.2R$, does 
the fraction increase quite suddenly to unity. To show this more quantitatively,
we present the ratio of the contribution to the overlap area from the corona
relative to the total area as a function of collision centrality in Au+Au
collisions in Fig.~\ref{rim}.
Here Fig.~\ref{rim}(a) gives the ratio, $R_w$, of 
the number of wounded nucleons in the corona to the total number 
of wounded nucleons while Fig.~\ref{rim}(b) gives the ratio, 
$R_c$, of the number of collisions in the corona relative to the total number 
of collisions. In both cases, the centrality dependence is shown as function 
of $N_{\rm part}/N_{\rm max}$, where $N_{\rm part}$ is the number of wounded 
nucleons and $N_{\rm max}=N_{\rm part}(b=0)$.

\medskip

\medskip

\begin{figure}[htb]
\centerline{\psfig{file=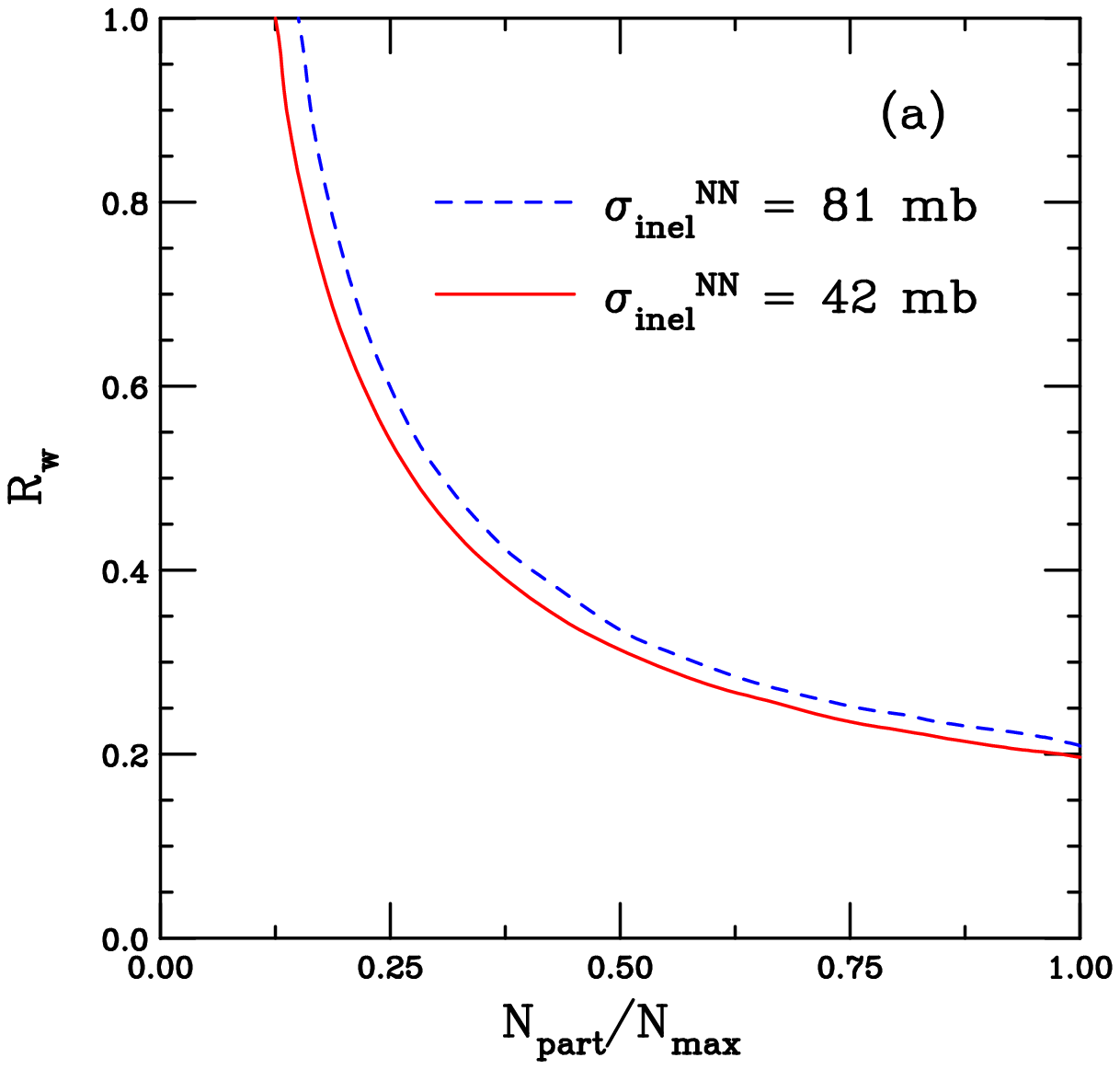,width=8cm}\hfill
\psfig{file=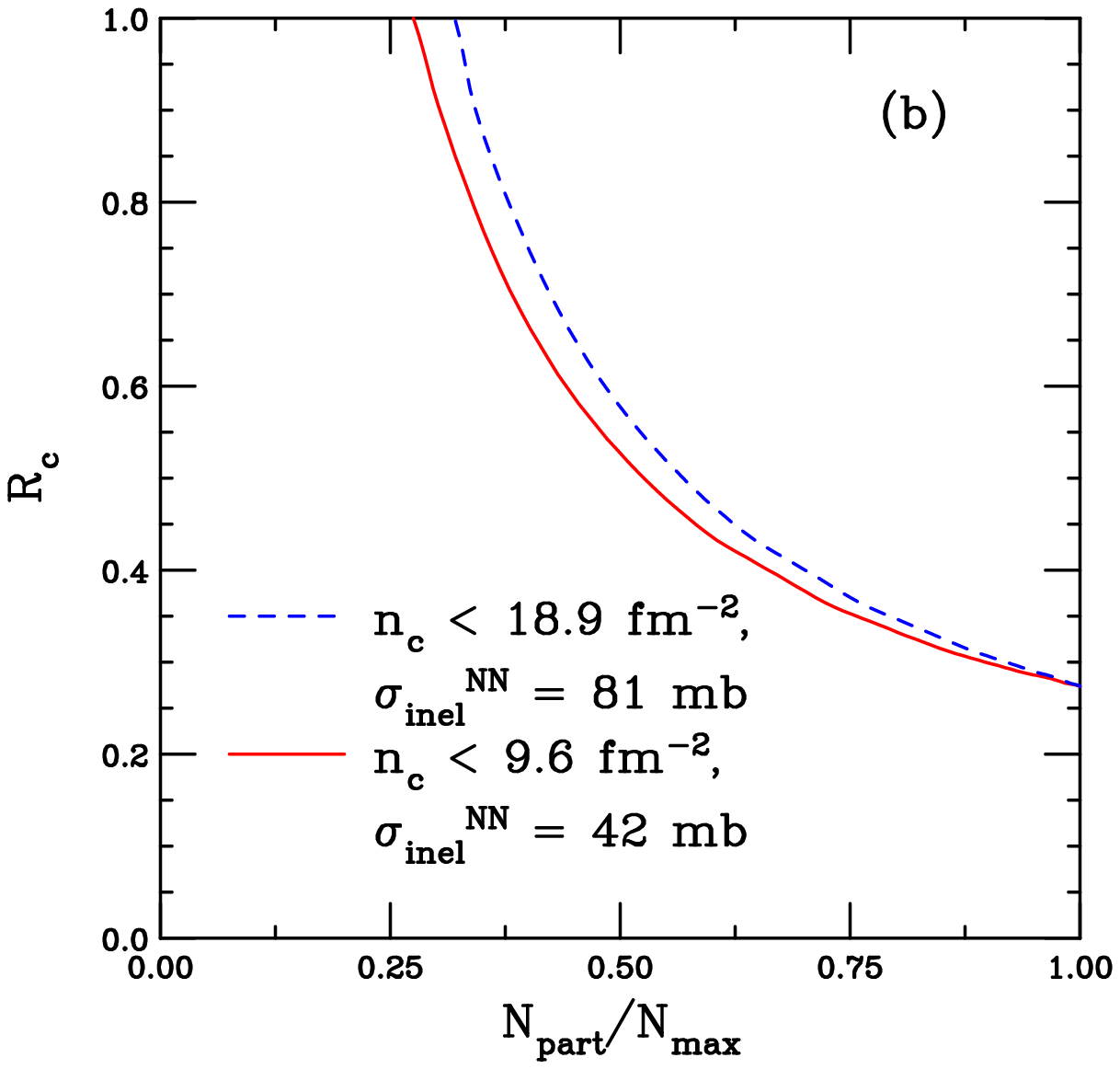,width=8cm}~~~~~~~}


\caption{Same as Fig.~\protect\ref{rim} except that now $\sigma_{\rm inel} =81$~mb
is also shown.}
\label{sigma}
\end{figure}

The results shown in Fig.~\ref{rim} employ
$\sigma^{NN}_{\rm inel} = 42$ mb, appropriate for $\sqrt{s_{NN}} = 200$ GeV 
collisions at RHIC.  For $AA$ collisions at the LHC, a larger cross section is 
more appropriate.  We choose $\sigma_{\rm inel}^{NN} = 81$~mb.  Increasing 
$\sigma^{NN}_{\rm inel}$ by nearly a factor of two
has only a relatively small effect 
on the behavior of $R_w$ and $R_c$, as shown in Fig.~\ref{sigma}. 
The value of $n_w$ is independent of $\sigma_{\rm inel}^{NN}$, see 
Eq.~(\ref{nwdef}).  However, the value  of $R$ at which $n_w(R)/n_w(0) < 0.5$ 
increases marginally, causing $R_w$ to shift to slightly higher values of
$N_{\rm part}/N_{\rm max}$ in Fig.~\ref{sigma}(a).
The origin of this shift is thus not due to $n_w(R)$ but to the increase in 
$N_{\rm part}$ overall since, at $\sigma_{\rm inel}^{NN}
\rightarrow \infty$, $N_{\rm max} \rightarrow 2A$ while finite values of 
$\sigma_{\rm inel}^{NN}$ reduce $N_{\rm max}$ relative to this asymptotic value.
Note that the value of $n_c$ defining the corona
is linearly dependent on $\sigma_{\rm inel}^{NN}$, see Eq.~(\ref{ncdef}), so that
for $\sigma_{\rm inel}^{NN} = 81$~mb, the value of $n_c = n_c(0)/2$ is 
18.9~fm$^{-2}$.  The value of $R$ when  $n_c = n_c(0)/2$ is independent of
$\sigma_{\rm inel}^{NN}$.  The same shift in $N_{\rm max}$ described above for $R_w$
manifests itself as a steepening of $R_c$ with $N_{\rm part}/N_{\rm max}$ for the
larger cross section.  Note that the value of $R_c$ for the two $NN$ cross
section values coincide for $N_{\rm part} = N_{\rm max}$.

\medskip

\medskip

\begin{figure}[htb]
\centerline{\psfig{file=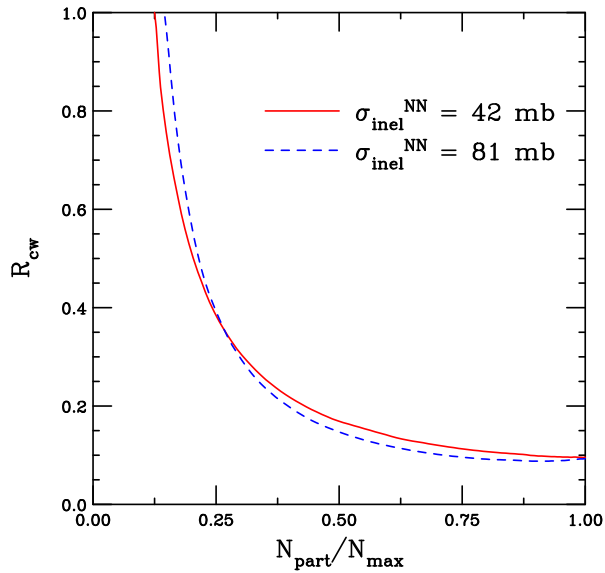,width=8cm}}

\caption{Ratio of collisions in the corona relative to the total number of 
collisions for Au+Au collisions, using $n_w$ to define the corona.
The result is shown for $\sigma^{NN}_{\rm inel} = 42$ and 81~mb.}
\label{w-c}
\end{figure}

The definition of the corona is clearly model dependent and can 
therefore be given in terms of either $n_w$ or $n_c$. However,
charmonium production is a hard process and, as such, its rate will be 
governed by the number of binary collisions. Excluding all other possible 
effects, Fig.~\ref{rim}(b) can therefore be directly compared to the centrality 
profiles of production data. If we define the corona in terms of wounded 
nucleons, we have to calculate the number of collisions in the corona and core
regions defined by $n_w$ in order to compare to the data. The
resulting ratio, $R_{cw}$, of the number of collisions in
the corona to the total number of collisions is shown in Fig.~\ref{w-c} for
both values of $\sigma_{\rm inel}^{NN}$. 

\medskip

We note that changing the collision system from Au+Au to Pb+Pb also has
only a small effect on $R_w$, $R_c$ and $R_{cw}$.  The central values of $n_w$ 
and $n_c$ will change somewhat.  In addition, the values of $R$ at which $n_w/
n_w(0) < 0.5$ and $n_c/n_c(0) < 0.5$ are slightly higher for the larger $A$.

\medskip

\medskip

Before any comparison to data, we have to address the role
of cold nuclear matter effects. In the usual treatments, where the
charmonium production in nuclear collisions is compared to that in
$pp$ or $pA$ interactions to determine ``anomalous'' behavior, the 
role of cold nuclear matter effects on charmonium production in both
the initial and final state has to be taken into account. 
If one considers instead, as suggested in Ref.~\cite{sridhar}, the ratio 
of charmonium to open charm, 
initial-state effects should largely cancel. In this context, a
first test would be to compare production at central and forward rapidities.
At RHIC, the $AA/pp$ ratios show differences for the two rapidities.
If the hidden to open charm ratios are measured as a function of centrality,
would the results in the different rapidity regions coincide? 
Final-state effects on the charmonium resonances will certainly be reduced 
in the corona.  However, 
it is not clear if they can be neglected entirely. In particular, 
at the SPS, there are indications that nuclear absorption of the produced 
charmonia in the nuclear medium is not negligible. Analysis of $J/\psi$
production at central rapidity over the range of fixed-target energies shows
that the effective absorption cross section decreases with $\sqrt{s_{NN}}$
\cite{LVW}.  This result is not unexpected since, as $\sqrt{s_{NN}}$ increases, 
the traversal time of the nascent charmonium state through the nuclear medium 
will become very short, diminishing 
the role of absorption. While absorption could still play a role in the 
comparison between central and forward rapidities, the 
hidden to open charm rates should clarify the situation considerably. 
 
\medskip

If we assume that at sufficiently high $\sqrt{s_{NN}}$, the central 
core of the transverse collision region is hot enough to dissolve all 
charmonium bound states, so that only those produced in the corona region
survive, then we obtain 
two immediate predictions for the ratios of charmonium to open charm 
production after the $B$-decay contributions have been removed.  
These predictions are the main result
of this paper.

\begin{itemize}
\item{The \J~survival probabilities in central $AA$ collisions for fixed $A$ 
should be essentially the same at the SPS, RHIC and the LHC.}
\item{The centrality profiles for \J~production at RHIC and the LHC should 
overlap.  Some differences may be possible for SPS energy if the
dissociation in the core is incomplete for more peripheral interactions.}
\end{itemize}.

\vskip-0.3cm
Here we have defined survival as the ratio of ratios: we divide the
charmonium to open charm production ratio in $AA$ collisions as 
function of centrality by the corresponding single-valued production ratio 
in $pp$ collisions. No scaling is needed, but if the $pp$ 
ratio is not independent of rapidity, the $AA$ and $pp$ ratios should be 
compared at the same rapidity. 

\medskip

We can make an additional prediction for the expected transverse momentum
distribution. Nuclear effects due
to initial-state scattering and final-state flow
generally broaden the $p_T$ distributions relative
to $pp$ interactions. If only corona production remains in high energy
$AA$ collisions, the $p_T$ distribution should
converge to that in $pp$ interactions. Such convergence would
discriminate against recombination models of \J~production which would predict
narrower $p_T$ distributions \cite{thews2}.  However, charm flow has
been observed in nuclear collisions \cite{charmflow} which, in the case of
statistical recombination, should be reflected as $p_T$ broadening
of the \J~produced by recombination. 

\bigskip

\centerline{\large \bf Acknowledgement}

\bigskip

The visit of S.D.\ at the University of Bielefeld was supported by IRTG 881.
The work of R.V.\ was performed under the auspices of the U.S.\ Department
of Energy by Lawrence Livermore National Laboratory und Contract
DE-AC52-07NA27344 and was also supported in part by the National 
Science Foundation Grant NSF PHY-0555660.

\end{document}